\documentstyle[preprint,revtex]{aps}

\tightenlines
\begin{document}
\begin{title}
Surface Effects on Bulk Plasmons
\end{title}
\author{Olle Heinonen$^1$ and Walter Kohn$^2$}
\begin{instit}
$^1$Department of Physics, University of Central Florida, Orlando, FL 32816\\
$^2$Department of Physics, and NSF Center for Quantized Electronic Structures\\
University of California, Santa Barbara, CA 93106
\end{instit}
\begin{abstract}
We discuss the effects of surfaces on bulk plasma oscillations in
a metallic film
%the collective perpendicular charge oscillations
%in a thin metallic slab
of thickness $L$. The leading
corrections to the plasma frequencies due to the surfaces
are proportional to $L^{-1}$.
%These corrections are due to the
%non-locality of the conductivity tensor $\sigma_{i,j}({\bf r};{\bf r}')$.
The frequencies of low-lying long wavelength modes are given by the
dispersion relation of bulk modes, $\omega=\omega_p[1+\alpha(q^2/k_F^2)]
$,
where the allowed $q$-values are $q^{(n)}=(n\pi/\overline L)$,
$\overline L\equiv L-2d_b$, and $d_b$ is a complex length characteristic of the
surface (reminiscent of but different from Feibelman's $d_\perp$).
An explicit expression for $d_b$ is derived. Resonant excitation
by an external field, $E_0e^{-i\omega t}$, is calculated.
\end{abstract}
\newpage

The widespread interest in nanostructures (thin films, layer structures,
quantum wires, quantum dots {\em etc.\/}) as well as in atomic and
molecular clusters, in  which at least one dimension  is ``mesoscopic''
(10 {\AA} -- 1000 {\AA}), has led us to re-examine the effects of
finite size on charge collective modes (plasmons). In an infinite
system, these modes have frequencies given by the
bulk plasmon relation \cite{Ichimaru}
\begin{equation}
\omega(q)=\omega_p+\alpha\left({q\over k_F}\right)^2\omega_p,\label{dispersion}
\end{equation}
where $\omega_p$ is the classical plasma frequency, $\omega_p^2=4\pi e^2 n/m$,
with $n$ the electron density, $k_F$ the Fermi wavevector, and
$\alpha$ a (complex) number. The imaginary part of $\alpha$ describes
the damping of bulk plasmons, which in an infinite system is
due to excitations of multiple
electron-hole pairs. In a finite system, we expect the following
qualitative changes in the plasmon spectrum due to the
presence of surfaces. First, only a discrete
set of frequencies will be allowed. Second, there will be
additional damping due to the decay of plasmons into (single) electron-hole
pairs. In this Letter, we will show how both the real and imaginary
parts of the allowed frequencies are completely determined by a
single (complex) length. Specifically,
we consider a thin slab with a positive jellium charge
background, $n_+=\overline n$ between $z=0$ and $z=L$ and
infinite in the $x$ and $y$ directions, and a
neutralizing electron liquid described by the equilibrium density
distribution $n_-(z)$ (See Fig. \ref{fign}). We shall discuss bulk-like
collective modes of the form
\begin{equation}
n({\bf r},t)=n(z)e^{-i\omega^{(n)} t},\label{bulkform}\end{equation}
where the eigenfrequencies $\omega^{(n)}$ are near the classical
bulk plasma frequency associated with the density $\overline n$,
$\omega_p=\left(4\pi e^2\overline n/m\right)^{1/2}$.

This problem has been previously addressed experimentally \cite{Anderegg,Ynsun}
and theoretically \cite{Feibelman1} for modes of the form (\ref{bulkform})
and of the more general form
\begin{equation}
n({\bf r},t)=n(z)e^{i(px-\omega t)}.\label{general}
\end{equation}
In the present Letter, limited to the case
$p=0$, we derive the following new result: the
(complex) eigenfrequencies $\omega^{(n)}$, near $\omega_p$, are
determined by the bulk plasmon dispersion Eq. (\ref{dispersion})
%\begin{equation}
%\omega(q)=\omega_p\left(1+\alpha(q/k_F)^2\right)\label{dispersion}
%\end{equation}
for small $q$, and by an effective, complex, slab thickness
\begin{equation}
\overline L=L-2d_b,\label{effectiveL}
\end{equation}
where $d_b$, precisely defined below, is an effective, complex,
surface thickness. The thickness $d_b$ is reminiscent of the parameter
$d_\perp$ introduced by Lang and Kohn \cite{Lang} for
$\omega=0$ and generalized for arbitrary $\omega$ by Feibelman
\cite{Feibelman2} but is {\em not} the same quantity. The allowed
(complex) interior wavenumbers, $q^{(n)}$, are given by
\begin{equation}
q^{(n)}\overline L=n\pi\quad(n=1,2,\ldots)\label{q}
\end{equation}
and the corresponding (complex) eigenfrequencies are given by Eq.
(\ref{dispersion}):
\begin{equation}
\omega^{(n)}=\omega(q^{(n)}).\label{eigen}\end{equation}

We now derive the above-stated results. We take the equilibrium
electron density,
$n_-(z)$, to vanish for $z<-c$ and $z>L+c$, and introduce an auxiliary
thickness \cite{independent} $a$, large compared to the surface
thickness but much smaller than $L$. We then distinguish two surface regions,
$-c<z<a$ and $L-a<z<L+c$, and the bulk region $a<z<L-a$, in
which $n_-(z)=\overline n$ (see Fig. \ref{fign}).

We consider a collective mode of frequency $\omega$, with self-consistent
electric field, $E(z)$, parallel to $z$. This field induces
a current density,
\begin{equation}
j(z)=\int_{-c}^{L+c}dz'\,\sigma(z,z')E(z'), \label{j}
\end{equation}
where
%\begin{equation}
$\sigma(z,z')=\int dx'\,dy'\,\sigma_{zz}(x-x',y-y';z,z')$;
%\label{sigma}\end{equation}
$\sigma_{zz}$ is the $zz$-component of the non-local
conductivity tensor. (For ease of notation, the time dependence,
$e^{-i\omega t}$, and the
functional dependence of fields, currents {\em etc.} on $\omega$ will
generally not be explicitly written out.)
The quantity $\sigma(z,z')$ is short-ranged
in the difference variable $(z-z')$; also, when both $z$ and $z'$ are in the
bulk region,
\begin{equation}
\sigma(z,z')=\sigma(z-z'),\quad a<z,z'<L-a\label{sigmabulk}
\end{equation}
equal to the bulk conductivity corresponding to $\overline n$ and $\omega$.
Of course $\sigma(z,z')$ vanishes when either $z$ or $z'$ are outside
the slab $(-c,L+c)$, as do $E(z)$ (due to charge neutrality) and $j(z)$.

Denoting the induced charge density by $n(z)$, the continuity equation
and Gauss' law are
\begin{equation}
i\omega e n(z)+{d\, j(z)\over dz}=0\label{continuity}
\end{equation}
and
\begin{equation}
{d E(z)\over dz}=-4\pi en(z)\label{Gauss}\end{equation}
(where $e$ has been taken as positive). Since at $z=-c$ both $j(z)$ and
$E(z)$ vanish, (\ref{continuity}) and (\ref{Gauss}) imply that, for
{\em all} $z$,
\begin{equation}
j(z)={i\omega\over 4\pi}E(z).\label{jE}
\end{equation}
We can eliminate $j(z)$ from (\ref{j}) and (\ref{jE}) which results in the
integral equation
\begin{equation}
{i\omega\over4\pi}E(z)=\int_{-c}^{L+c}dz'\,\sigma(z,z')E(z').\label{integral}
\end{equation}

We are looking for solutions, which, in the bulk, behave as
$e^{\pm iqz}$, with $q\ll k_F$. In this limit of long
wavelengths in the bulk, we can, with Eq. (\ref{sigmabulk}) in
Eq. (\ref{integral}) obtain the frequencies of these solutions by expanding
$E(z')$ about $z$ when $z$ and $z'$ are in the bulk. The result is
\begin{equation}
{i\omega\over 4\pi}=\sigma_0(\omega)-\frac{1}{2}\sigma_2(\omega)q^2,\label{wq}
\end{equation}
where
\begin{eqnarray}
\sigma_0(\omega)&=&\int dz'\,\sigma(z-z';\omega)={i\omega_p^2\over 4\pi\omega},
\label{sigma0}\\
\sigma_2(\omega)&=&\int dz'(z'-z)^2\sigma(z-z';\omega).\label{sigma2}
\end{eqnarray}
Equation (\ref{wq}) is the bulk plasmon dispersion relation of
an infinite system of conductivity $\sigma(z-z')$.
In the limit $L\gg a$, the frequencies of the lowest-lying eigenmodes of
Eq. (\ref{integral}) are very near $\omega_p$ and the shape of $E(z)$
in the surface region tends to the limiting shape
\begin{equation}
E(z)\rightarrow \widetilde E(z)\equiv E(z;\omega_p)\label{tildeE}
\end{equation}
shown schematically in Fig. \ref{figE}. We arbitrarily take
$E(a)=1$ and write, near $z=a$
\begin{equation}
\widetilde E(z)={z-d_b\over a-d_b}\label{Esurf}\end{equation}
where $d_b$ is the, as yet undetermined, point where $\widetilde E(z)$
extrapolates to
0. When $z$ is in the bulk the solution which joins (\ref{Esurf}) near
$z=a$ is
\begin{equation}
E(z)={1\over q(a-d_b)}\sin q(z-d_b).\label{sinE}\end{equation}
Thus, $d_b$ plays the role of a (complex) scattering length.

The
length $d_b$ can be simply expressed in terms of the limiting
surface density distribution, $\widetilde n(z)$, associated
with collective modes
with $q\rightarrow0$ and $\omega\rightarrow\omega_p$. From Eq. (\ref{Gauss})
and the form of $E(z)$ in the surface region we see that $\widetilde n(z)$
has the
general form shown in Fig. \ref{figE}. Thus, for $z$ near $a$, we have
\begin{eqnarray}
{\widetilde E(z)\over(-4\pi e)}&=&\int_{-c}^z dz'\,\widetilde n(z)=
\int_0^z dz'\,
\left[{d\over dz'}\left(z'\widetilde n(z')\right)-z'
{d\widetilde n(z')\over dz'}\right]\nonumber\\
&=&zn_{-c}-\int_{-c}^\infty dz'\,z'{d\widetilde n(z')\over dz'}\nonumber\\
&=&z\int_{-c}^\infty dz'\,{d\widetilde n(z')\over dz'}-\int_{-c}^\infty dz'\,
z'{d\widetilde n(z')\over dz'}.\label{newE}\end{eqnarray}
The upper limits $\infty$ are to be understood as $\gg k_F^{-1}$
but $\ll L$. The linear form of $E(z)$ near $a$, Eq. (\ref{newE}),
extrapolates to zero when $z$ equals
\begin{equation}
d_b\equiv{\int_{-c}^\infty dz\,z{d\widetilde n(z)\over dz}\over
\int_{-c}^\infty dz\,{d\widetilde n(z)\over dz}}.\label{defdb}\end{equation}
Thus, $d_b$ is the center of mass of the function $d\widetilde n(z)/dz$.

The eigenmodes are either even or odd about $L/2$ so that, by Eq. (\ref{sinE}),
the allowed values of $q$ are
\begin{equation}
q^{(n)}\overline L=n\pi,\quad (n=1,2,\ldots)\end{equation}
where $\overline L$ is give by Eq. (\ref{effectiveL}). The corresponding
frequencies are given by Eq. (\ref{dispersion}). Using the notation
$d_b=d_b'+id_b''$ {\em etc\/}, we have
\begin{eqnarray}
{\omega^{(n)}}'&=&\omega_p+{1\over k_F^2}\left(\alpha'-4\alpha''{d_b''\over
\overline L'}\right)
\left({n\pi\over\overline L'}\right)^2
\label{oprime}\\
{\omega^{(n)}}''&=&{1\over k_F^2}\left(\alpha''+4\alpha'{d_b''\over
\overline L'}\right)\left({n\pi\over\overline L'}\right)^2,
\label{otwoprime}
\end{eqnarray}
where $\overline L'=L-2 d_b'$. We observe that the real parts Eq.
(\ref{oprime})
of the frequencies correspond to bulk plasmon modes with the boundary
conditions $E(d_b')=E(L-d_b')=0$. The imaginary parts, (\ref{otwoprime}),
have contributions from the bulk ($\propto\alpha''$) and from the
surface ($\propto(d_b''/\overline L')$).

We have solved the integral equation (\ref{integral}) in the discrete
form
\begin{equation}
{i\omega\over 4\pi}E_m=b\sum_{n=0}^N\sigma_{mn}E_n,\label{discrete}
\end{equation}
where $b\equiv L/N$ is the interval length. We have chosen
$L=35.4$ {\AA}, $N=248$, $b=0.14$ {\AA}, and $c=0$.
% and a conductivity given by
%\begin{eqnarray}
%\sigma_{m,m}&=&i\mu(m){1\over 1+e^{-3(m-3)/2}
%-0.1\Im\left[\mu(m)\]{k\over\left[1+e^{-3(m-3)/2\right]^2}
%\nonumber\\
%\sigma_{m,m\pm1}&=&i\nu(m){1\over 1+e^{-3(m-3)/2}
%-0.1\Im\left[\nu(m)\]{k\over\left[1+e^{-3(m-3)/2\right]^2}\label{25}
%\end{eqnarray}
%and symmetrically for $m>51$. In the bulk ($m\agt4$), the conductivity is
%$\mu(\omega)=7.273\times10^{-2}-i1.967\times10^{-3}, and
%$\nu(\omega)=-2.302\times10^{-2}+i9.836\times10^{-4}.
%The conductivity given by (\ref{25}) corresponds
%to $k_F=1.75\times10^8$ cm$^{-1}$, $\hbar\omega_p=15.8$ eV, and
%$\alpha=0.585+i0.025$ (appropriate for Al), in a slab with the
%the jellium edge is
%located at $z=3b$ and a surface thickness characterized by a length
%$(2/3)b$. The small real part was added to the conductivity in the
%surface regions to model decay of the plasma eigenmodes into
%electron-hole pairs.
For the sake of illustration, we have chosen the
following values for $\sigma_{mn}$. In the ``bulk'', $4\leq m\leq 125$, we
have chosen
\begin{eqnarray}
\sigma_{m,m}=i\mu(\omega)={i\omega_p^2\over4\pi\omega b}
-{\omega_p\over3\pi k_F^2 b^3}\left[\alpha_2-i\alpha_1\right],&\quad&
\sigma_{m,m+2}=\sigma_{m,m-2}=\frac{1}{2}i\nu\nonumber\\
\sigma_{m,m+1}=\sigma_{m+1,m}=i\nu=
{2\omega_p\over21\pi k_F^2b^3}\left[\alpha_2-i\alpha_1\right],&\quad&
\sigma_{m,m+3}=\sigma_{m,m-3}=\frac{1}{4}i\nu
\label{discsigma}
\end{eqnarray}
and symmetrically for $m>125$, with $k_F=1.75\times10^8$
cm$^{-1}$, $\hbar\omega_p=15$ eV, and
$\alpha=0.57+i0.025$ in Eq. (\ref{dispersion}) (appropriate
\cite{Ichimaru} for Al in the limit of $q/k_F\ll1$). The functions
$\mu(\omega)$
and $\nu$ are chosen so that the average and second moment of
$\sigma_{mn}$ are $\sigma_0(\omega)$ and $\sigma_2(\omega_p)$. In the
``surface'',  $0\leq m\leq3$, we have chosen
\begin{eqnarray}
\sigma_{0n}&=&\sigma_{n0}=0\nonumber\\
\sigma_{m,m}&=&\frac{m}{4}\left(i\mu(\omega)+\epsilon\right)\nonumber\\
%% FOLLOWING LINE CANNOT BE BROKEN BEFORE 80 CHAR
\sigma_{m,m+1}&=&\sigma_{m+1,m}=\frac{m}{4}\left(i\nu+\epsilon\right)\nonumber\\
%% FOLLOWING LINE CANNOT BE BROKEN BEFORE 80 CHAR
\sigma_{m,m+2}&=&\sigma_{m+2,m}=\frac{m}{8}\left(i\nu+\epsilon\right)\nonumber\\
\sigma_{m,m+3}&=&\sigma_{m+3,m}=\frac{m}{16}\left(i\nu+\epsilon\right)
\label{surfsigma}
\end{eqnarray}
and symmetrically near the other surfaces. The small real part
$\epsilon=\omega_p\alpha''/(6\pi k_F^2b^3)$
was added to the conductivity in the surface regions as
a simple model of damping in the surfaces
due to decay into electron-hole pairs.
Apart from this small real part, equation (\ref{surfsigma})
is a simple interpolation between $\sigma_{0n}=\sigma_{n0}=0$ and
the bulk conductivity. The results for the lowest three modes are plotted
in Figs. \ref{numfig} and \ref{numfig2}, and confirm our conclusions, Eqs.
(\ref{q}), (\ref{eigen}), (\ref{oprime})
and (\ref{otwoprime}).

How do our results relate to charge density waves in classical Maxwell theory?
``Classical'' here means a local $\sigma(z,z')$
\begin{equation}
\sigma_{\rm class}(z,z')=\sigma_0\delta(z-z')={i\omega_p^2\over4\pi\omega}
\delta(z-z'),\label{classical}\end{equation}
so that (\ref{integral}) becomes
\begin{equation}
{i\omega\over4\pi}E(z)={i\omega_p^2\over4\pi\omega}E(z).\label{classE}
\end{equation}
The boundary conditions \cite{wrongBC} are, from charge neutrality,
%\begin{equation}
$E(0)=E(L)=0$.
%\label{classBC}\end{equation}
Thus, classically the general solution is $\omega=\omega_p$ and
%\begin{equation}
$E(z)=F(z)$,
%\label{arbitrary}\end{equation}
an {\em arbitrary} function of $z$ satisfying the boundary conditions
$E(0)=E(L)=0$.
%The corresponding density is
%\begin{equation}
%n(z)=-{1\over4\pi e}{dF\over dz}.\label{dF}\end{equation}
In particular, the functions
\begin{eqnarray}
F_n(z)&=&\sin(q^{(n)}_{\rm class}z)\quad q_{\rm class}^{(n)}\equiv{n\pi\over
L},
\;n=0,1,\ldots\\
\widetilde\omega^{(n)}&\equiv&\omega_p,\quad \mbox{all $n$,}\label{tut}
\end{eqnarray}
are solutions. Thus, our {\em non-local} and translationally
{\em non-invariant} $\sigma(z,z')$ leads to:
(1) replacement of $L$ by the complex $\overline L=L-2d_b$; (2) a {\em finite}
dispersion of $\omega^{(n)}$; and, if $\sigma(z,z')$ is known in detail
in the surface region, the detailed field and density distributions
in the surface region.

We now turn to resonant response to a uniform external field,
$E_0e^{-i\omega t}$. Eqs. (\ref{j}), (\ref{continuity}) and (\ref{Gauss})
remain unchanged, but in Eq. (\ref{jE}) $E(z)$ is replaced by $E(z)-E_0$
and Eq. (\ref{integral}) becomes
\begin{equation}
{i\omega\over 4\pi}\left[E(z)-E_0\right]
=\int_{-c}^{L+c}\,dz'\,\sigma(z,z';\omega)E(z').\label{34}\end{equation}
Near $\omega=\omega_p$ we can write this as
\begin{equation}
\left[{\omega_p\over4\pi}E(z)-\int dz'{\sigma(z,z')\over i}E(z')\right]
+\eta\left[{1\over4\pi}E(z)-\int dz'{\tau(z,z')\over i}E(z')\right]
={(\omega_p+\eta)\over4\pi}E_0,\label{35}\end{equation}
where $\sigma(z,z')\equiv\sigma(z,z';\omega_p)$, $\tau(z,z')\equiv\left[
d\sigma(z,z';\omega)/d\omega\right]_{\omega=\omega_p}$ and
$\eta\equiv\omega-\omega_p$. On the right-hand side, $\eta$ can be neglected.

Setting $E_0=0$, for a moment, gives us the equation for the slightly
complex normal modes, $E_n(z)$, and eigenvalues, $\eta_n=\eta_n'+i\eta_n''$.
The kernels $\sigma(z,z')/i$, $\tau(z,z')/i$ are symmetric and (slightly)
complex. The $E_n(z)$ satisfy the orthonormality relations
\begin{equation}
\int_{-c}^{L+c}\int_{-c}^{L+c}dz\,dz'\,E_m(z)\left[{1\over4\pi}\delta(z-z')
-{\tau(z,z')\over i}\right]E_n(z')=\delta_{mn}.\label{ortho}\end{equation}
We now substitute the Ansatz
\begin{equation}
E(z)=\sum_n A_nE_n(z)\label{ansatz}\end{equation}
into (\ref{35}), giving
\begin{equation}
\sum_nA_n(\eta-\eta_n)E_n(z)={\omega_p\over4\pi}E_0.\label{38}\end{equation}
Using the orthonormality relation Eq. (\ref{ortho}) leads to
\begin{equation}
A_n={\omega_pE_0\over4\pi(\eta-\eta_n)}
\int_{-c}^{L+c}dz\int_{-c}^{L+c}dz'E_n(z')
\left[{1\over4\pi}\delta(z-z')-{\tau(z,z')\over i}\right].\label{39}
\end{equation}
Only even modes, $n=2,4\ldots$ are excited. A well defined resonance
occurs for $\eta\approx\eta_n'$ if
$\eta_n''\ll|\eta_n'-\eta_{n\pm2}'|$. In this case as single term in
(\ref{ansatz})
is dominant, giving
\begin{equation}
E_n(z)=E_0{\omega_p\left[(\eta-\eta_n')-i\eta_n''\right]\over
4\pi\left[(\eta-\eta_n')^2+{\eta_n''}^2\right]}.\label{40}\end{equation}
In order to resolve the discrete modes experimentally, the
spacing between the modes must be larger than the width.
{}From Eqs. (\ref{oprime}) and (\ref{otwoprime}) we find that
the spacing and width are given by
\begin{eqnarray}
\left(\eta_n'-\eta_{n-2}'\right)&=&
{\alpha'\over k_F^2}\left({\pi\over\overline L'}\right)^2
(4n-4)\label{41}\\
\eta_n''&=&{1\over k_F^2}\left(\alpha''+{4\alpha'd_b''\over\overline L'}\right)
\left({\pi\over\overline L'}\right)^2n^2.\label{42}\end{eqnarray}
Thus, as $n$ increases, the width gains on the spacing. For a film of
width $L'=30$ {\AA}, $d_b'=d_b''=1$ {\AA}, and
$\alpha'=0.57$, $\alpha''=0.025$ (Al), the highest well defined resonance is
$n\approx20$.

We plan to address the considerably more complex theory of modes with
finite wavevector $p$ (Eq. \ref{general}) in the near future.

We thank
Professor G.C. Strinati for discussions.
Support by the Office of Naval Research,
Grant No. ONR-N00014-89-J-1530 and the National Science Foundation
Grant No. NSF-DMR-90-01502 and NSF Science and Technology Center for Quantized
Electronic Structures DMR91-20007 is gratefully acknowledged.

\figure{Positive background density, $n_+(z)$ (dashed line), and
equilibrium electron density, $n_-(z)$ (dotted line).\label{fign}}
\figure{The limiting ($\omega\rightarrow\omega_p$) form, $\widetilde E(z)$,
and mode density $\widetilde n(z)$ in the surface regions. The mode
density $\widetilde n(z)$ approaches the constant value $n_0$ in the
bulk.\label{figE}}
\figure{(a) The real part of the electric field
of the three lowest bulk-like eigenmodes near the surface. In the surface,
the fields
all approach the limiting form $\widetilde E(z)$. {\em Insert} The real
parts of the three lowest bulk-like eigenmodes
for $0<z<L/2$. In the bulk region, the fields are sinusoidal. The vertical
dashed line marks the position of $a$.
\label{numfig}}
\figure{ (a) Real and (b) imaginary parts of the plasmon eigenfrequencies.
The straight lines are the dispersions Eqs. (\ref{oprime}) and
(\ref{otwoprime}) with $L'=34.8$ {\AA} and ${d_b}''=6\times10^{-2}$ {\AA}
obtained from Eq. (\ref{defdb}).\label{numfig2}}


\begin{references}
\bibitem{independent} Our final result is independent of the exact choice
of $a$ and $c$.
\end{references}
\end{document}